\documentclass{moriond}

\def\be{\begin{equation}}
\def\ee{\end{equation}}
\def\bea{\begin{eqnarray}}
\def\eea{\end{eqnarray}}

\newcommand{\Photo}{\includegraphics[height=35mm]{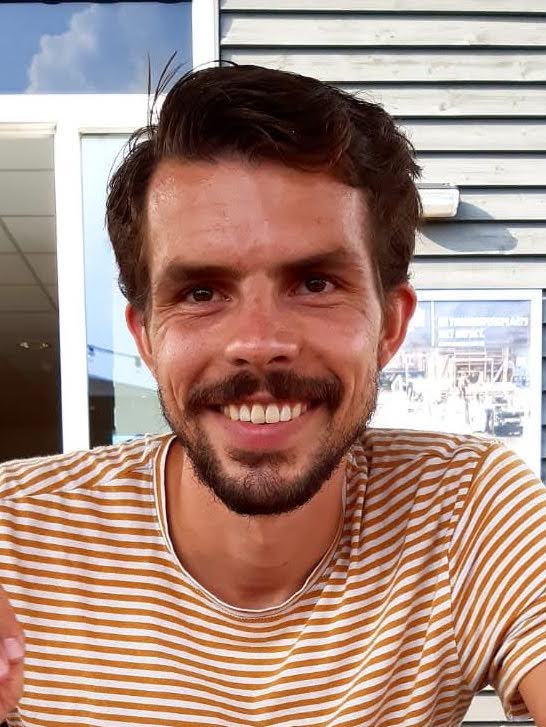}}

\begin{document}
\vspace*{4cm}
\title{RECENT ALICE RESULTS ON CHARM PRODUCTION AND HADRONISATION}
\author{L.A. VERMUNT, ON BEHALF OF THE ALICE COLLABORATION}

\address{Institute for Subatomic Physics, Utrecht University}

\maketitle
\abstracts{
Studies on the production of open charm hadrons are of paramount importance to investigate the charm-quark hadronisation mechanisms at the LHC, particularly through the evolution of the production ratio between different charm-hadron species. Measurements performed in pp and p--Pb collisions at the LHC have revealed unexpected features, qualitatively similar to what observed in larger systems and, in the charm sector, not in line with the expectations based on previous measurements from $\rm e^{+}e^{-}$ colliders and in ep collisions. These results suggest that charm fragmentation fractions might not be universal and that the baryon-to-meson ratio depends on the collision system. Model calculations that better reproduce the $\Lambda_{\rm c}^{+}/{\rm D}^{0}$ ratio in pp collisions expect a significant contribution to $\Lambda_{\rm c}^{+}$ yield from decays of heavier charm-baryon states, rely on hadronisation via recombination mechanisms, or are based on new colour reconnection topologies. In this contribution, the most recent results on open heavy-flavour production in pp and Pb--Pb collisions measured by the ALICE Collaboration will be discussed. Emphasis will be given to the discussion of the impact of these studies on our understanding of the hadronisation processes.
}

\section{Introduction}

Heavy quarks (charm and beauty) are predominantly produced in the initial stages of the collisions via hard-scattering processes and provide a sensitive test of perturbative quantum chromodynamics (pQCD) calculations. These predictions utilise the QCD factorisation approach, calculating the transverse momentum ($p_{\rm T}$) differential production as a convolution of (i) the parton distribution functions, (ii) the partonic cross section, and (iii) the fragmentation functions. This latter term parametrises the (non-perturbative) hadronisation of a heavy quark into a heavy-flavour hadron, which is usually assumed to be universal among different collision systems and energies and therefore tuned on $\rm e^{+}e^{-}$ and ep collision data. By comparing the production of different species of heavy-flavour hadrons, this assumption, and hadronisation of heavy quarks in general, can be investigated. With LHC Run 1 and 2 data, it has been shown that such pQCD calculations generally describe the $\rm D$- and $\rm B$-meson sector within uncertainties~\cite{Andronic:2015wma}, where instead heavy-flavour baryon production in pp collisions is less well understood~\cite{Acharya:2017kfy}.

In the presence of a quark--gluon plasma (QGP), the strongly-interacting colour-deconfined state of matter created in ultra-relativistic heavy-ion collisions, an additional hadronisation mechanism known as recombination (or coalescence) is considered. Here, soft quarks from the medium recombine with the heavy quark to form a meson or baryon. By studying heavy-flavour production in nucleus--nucleus collisions, the relevance of recombination in the medium for heavy quarks can be probed. Models that include hadronisation via recombination for charm quarks were observed to qualitatively describe $\rm D$-meson measurements in A--A collisions at RHIC and LHC~\cite{Andronic:2015wma}. The possibility of recombination in pp collisions is one of the topics of investigation to provide a better description of heavy-baryon production.

In ALICE, open charm hadrons are measured at midrapidity ($|y|<0.5$) via the decay channels $\rm D^{0} \to K^{-}\pi^{+}$, $\rm D^+ \to K^- \pi^+ \pi^+$, and $\rm \Lambda_{\rm c}^{+}$ to either $\rm pK^{-}\pi^{+}$ or $\rm pK^{0}_{\rm S}$. Heavier charm-baryon states are recently measured as well, exploiting the decay channels $\rm \Sigma_c^{0,++} \to \pi^{-,+} \Lambda_{\rm c}^{+}$, $\rm \Xi_c^0 \to \Xi^-\pi^+$ and $\Xi^- e^+ \nu_e$, and $\rm \Xi_c^+ \to \Xi^- \pi^+ \pi^+$. For the hadronic decay channels, invariant-mass fits are used to extract the charm-hadron raw yields after having applied selections on the displaced decay-topologies and on the particle-identification information of the daughter tracks to improve the statistical significance of the signal. The semi-leptonic decay channel for the $\rm \Xi_c^0$ baryon exploits instead a subtraction of same charge-sign from opposite charge-sign $\rm e\Xi$-pairs. The raw yields are corrected for the reconstruction and selection efficiency using Monte Carlo simulations and for the prompt fraction estimated using a FONLL-based approach~\cite{Acharya:2019mgn,Cacciari:2012ny}.

\section{Results}

\begin{figure}[tb!]
    \centering
    \includegraphics[width=0.4275\textwidth]{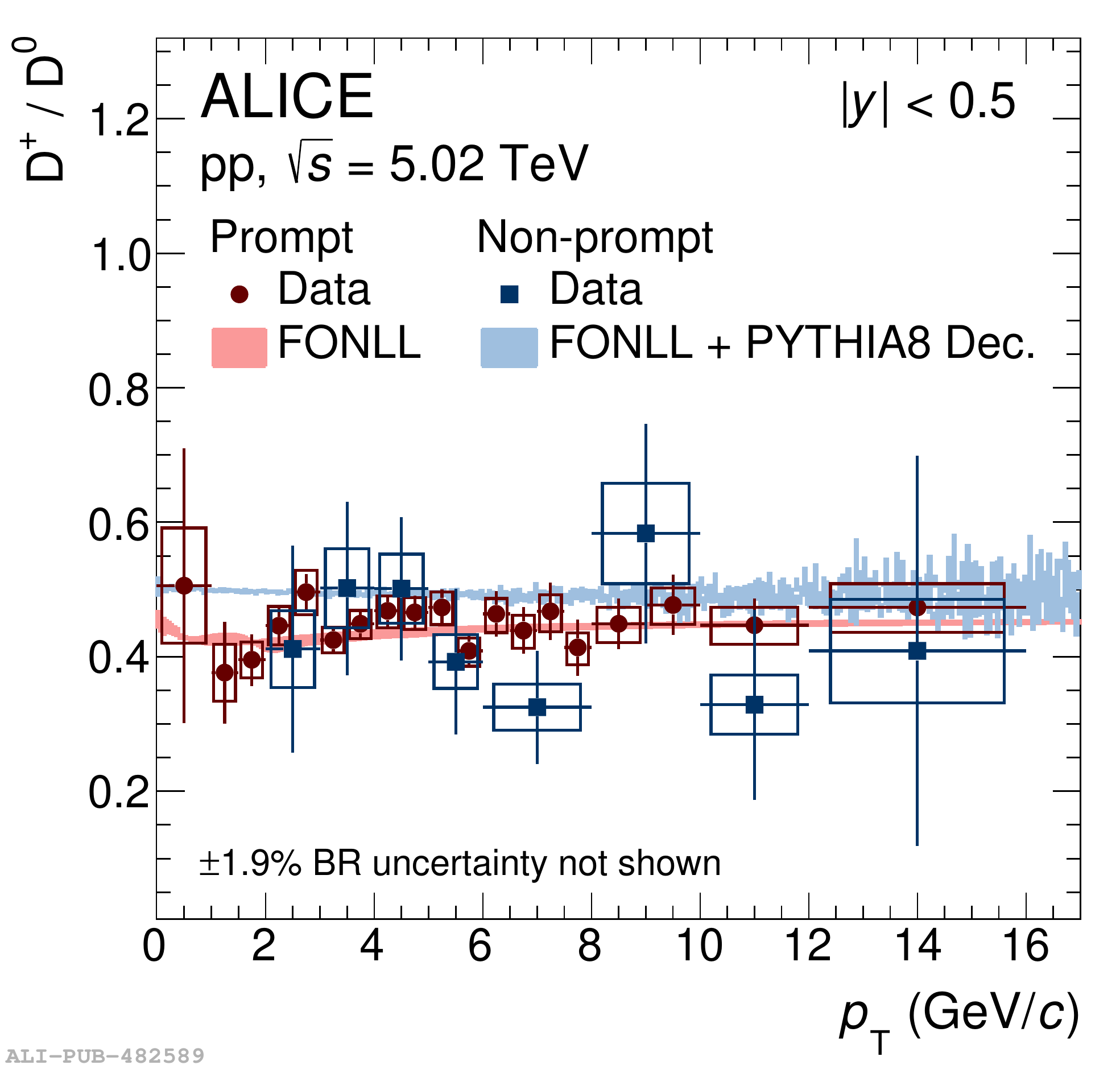}%
    \includegraphics[width=0.42275\textwidth]{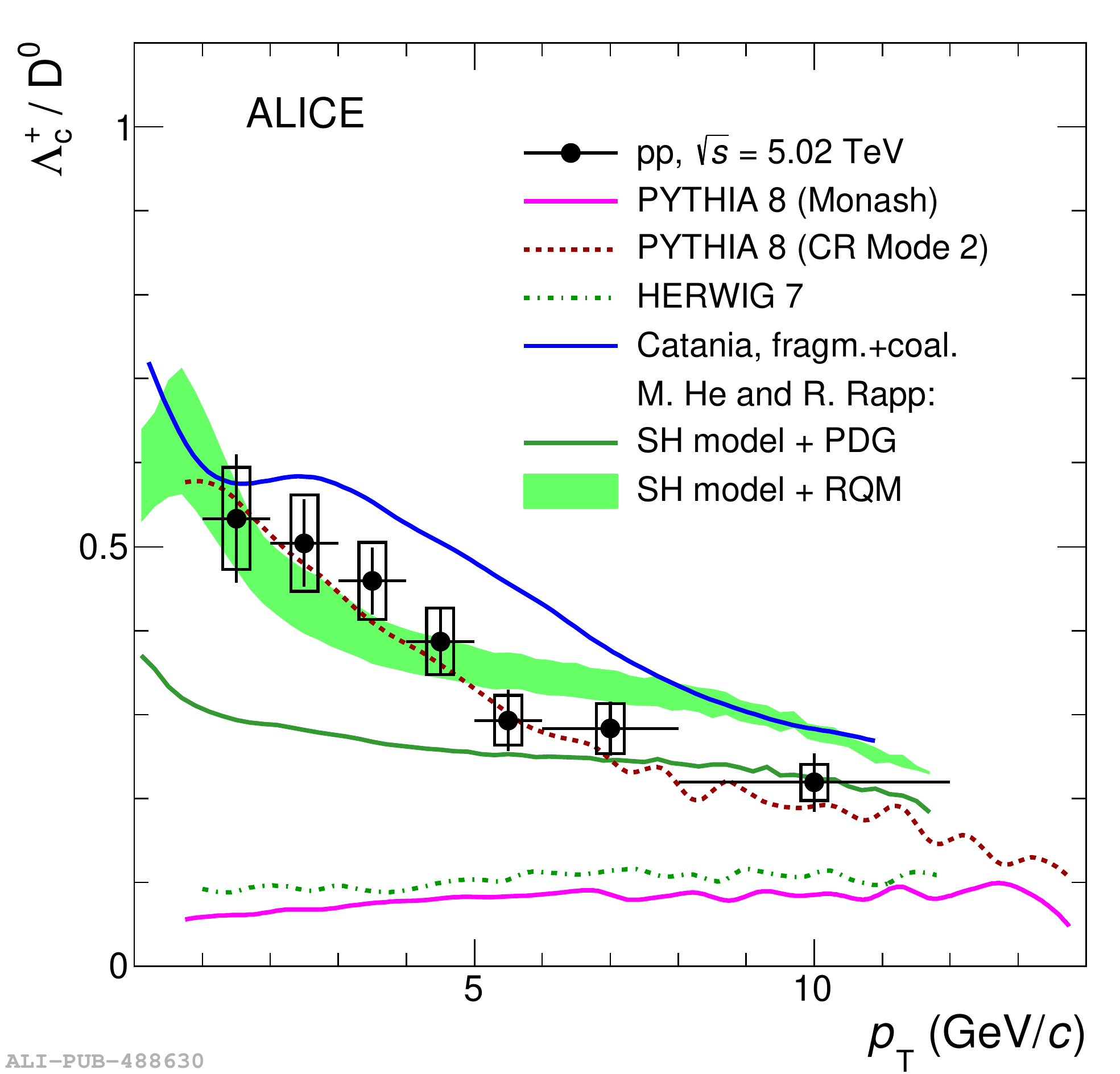}
    \caption{Left: Prompt and non-prompt $\rm D^+/D^0$ production ratios compared to pQCD predictions from FONLL calculations~\protect\cite{Cacciari:2012ny} as measured in $\sqrt{s} = 5.02$~TeV pp collisions~\protect\cite{Acharya:2021cqv}. Right: The $\Lambda_{\rm c}^{+}/{\rm D^0}$ ratio measured in pp collisions at 5.02~TeV~\protect\cite{Acharya:2020uqi} compared to different theoretical calculations~\protect\cite{Skands:2014pea,Bahr:2008pv,Christiansen:2015yqa,Minissale:2020bif,He:2019tik}.}
    \label{fig:charmhadronratios}
\end{figure}

The production of several charm hadrons is recently measured in minimum-bias pp collisions at $\sqrt{s} = 5.02$~TeV by ALICE~\cite{Acharya:2021cqv,Acharya:2020uqi}. The production ratios of different hadron species, shown in Fig.~\ref{fig:charmhadronratios} for $\rm D^+ / D^0$ and $\rm \Lambda_c^+ / D^0$ as function of the transverse momentum $p_{\rm T}$, can help to study heavy-flavour hadronisation. Both the prompt and non-prompt (coming from $\rm B$-hadron decays) ratio of non-strange $\rm D$ mesons are compatible with pQCD calculations using fragmentation fractions extracted from $\rm e^{+}e^{-}$ collision data~\cite{Cacciari:2012ny,Gladilin:2014tba}. Such predictions, however, significantly underpredict the measured $\rm \Lambda_c^+ / D^0$ ratio~\cite{Skands:2014pea,Bahr:2008pv}, indicating that the fragmentation fractions might not be universal. The baryon-to-meson ratio in pp collisions is better described by models with an extension of colour reconnection beyond the leading colour approximation~\cite{Christiansen:2015yqa}, models relying on hadronisation via recombination after the formation of a colour-deconfined state of matter~\cite{Minissale:2020bif}, or statistical hadronisation models with an augmented set of baryon states predicted by the relativistic quark model (RQM)~\cite{He:2019tik}. Nevertheless, the $\rm \Lambda_c^+$ puzzle in pp collisions remains an active research topic.

\begin{figure}[tb!]
    \centering
    \includegraphics[width=0.82\textwidth]{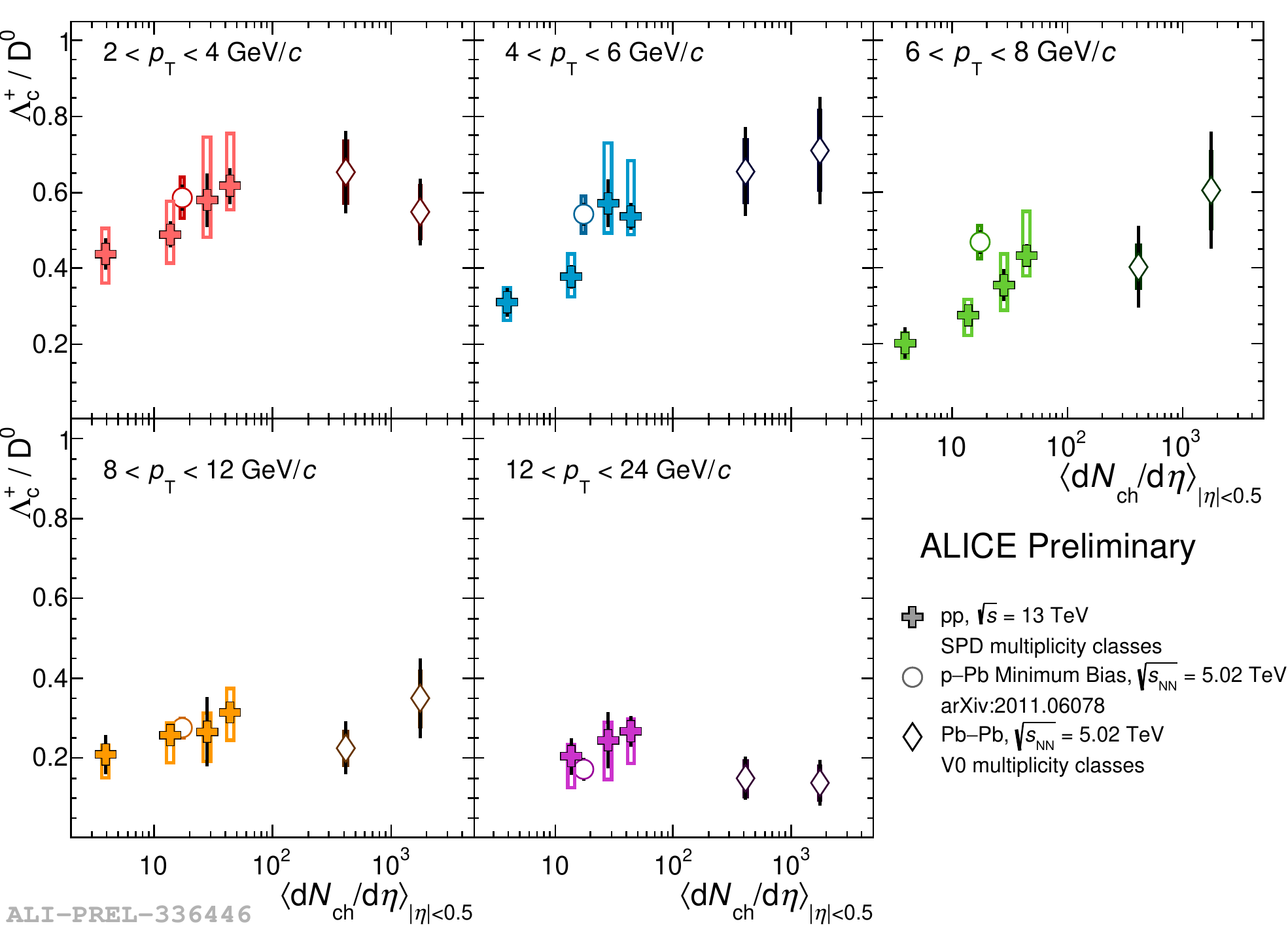}
    \caption{The $\Lambda_{\rm c}^{+}/{\rm D^0}$ production ratios, as a function of primary charged particles per unity of pseudorapidity, in pp, p--Pb, and Pb--Pb collisions at $\sqrt{s} = 13$~TeV and $\sqrt{s_{\rm NN}} = 5.02$~TeV, in five $p_{\rm T}$ intervals from 2 to 24~GeV/$c$.}
    \label{fig:LcD0_versus_mult}
\end{figure}

These observations triggered additional studies on charm-baryon production as function of event multiplicity. By comparing the $\rm \Lambda_c^+ / D^0$ ratio in high-multiplicity pp collisions with the values measured in Pb--Pb collisions, it can be investigated if recombination processes already start to play a role in pp collisions. Furthermore, low-multiplicity pp events might shed light on the differences observed among the different collision systems. In Fig.~\ref{fig:LcD0_versus_mult}, the $\rm \Lambda_c^+ / D^0$ ratio is presented as function of primary charged particles per unity of pseudorapidity. A rising trend with multiplicity is clearly observed at intermediate $p_{\rm T}$ ($4<p_{\rm T}<8$~GeV/$c$), where the baryon-to-meson ratios in the highest multiplicity pp events are compatible with measurements in Pb--Pb collisions. The gap with respect to the LEP average ($0.113\pm0.013{\rm(stat)}\pm0.006{\rm(syst)}$~\cite{Acharya:2017kfy}) is, however, not ``bridged'' when considering ultra-low multiplicity pp collisions. The multiplicity dependence in the baryon-to-meson ratio in pp systems is qualitatively reproduced by PYTHIA~8 calculations with enhanced colour reconnection~\cite{Christiansen:2015yqa}, although the observed compatibility in magnitude in high-multiplicity pp and Pb--Pb collisions may suggest that similar hadronisation mechanisms are at play as well. Considering Pb--Pb collisions, the hypothesis of a relevant contribution of charm recombination is further supported by recent measurements on the LHC Run 2 Pb--Pb data sample, especially for $\rm \Lambda_c^+$ and $\rm D^{+}_{s}$ production. 

\begin{figure}[tb!]
    \centering
    \includegraphics[width=0.9\textwidth]{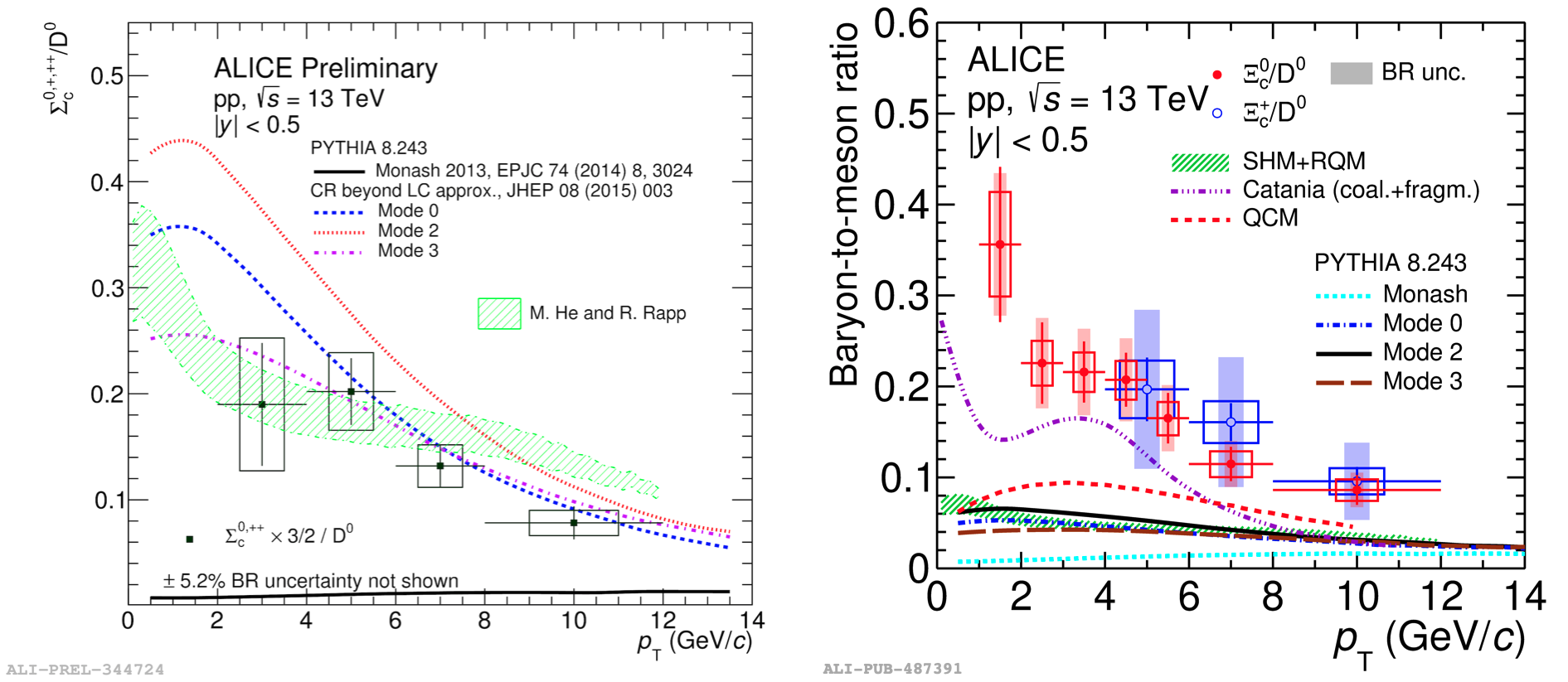}
    \caption{The $\rm \Sigma_c^{0,+,++}(2455)/D^0$ (left) and $\rm \Xi_c^{0,+}/D^0$ (right) production ratios as measured by ALICE in $\sqrt{s} = 13$~TeV pp collisions~\protect\cite{Acharya:2021vjp}. The measured ratios are compared to different theoretical calculations~\protect\cite{Skands:2014pea,Christiansen:2015yqa,Minissale:2020bif,He:2019tik}.}
    \label{fig:sigmac_xic}
\end{figure}

The ALICE Collaboration recently measured the production of heavier charm-baryon states in pp collisions at $\sqrt{s} = 13$~TeV~\cite{Acharya:2021vjp}. The measurement of ground-state $\rm \Sigma_c^{0,+,++}(2455)$ is a key ingredient to understand charm-baryon hadronisation, since a sizeable contribution to the $\rm \Lambda_c^+$ production via strong decays is expected. Measurements of the heavier charm-strange baryons $\rm \Xi_c^0$ and $\rm \Xi_c^+$ will pose further important constraints to charm-quark hadronisation models and, in addition, are a mandatory contribution for an accurate measurement of the $\rm c\overline{c}$ cross section~\cite{Acharya:2021set}. The ratios of the production of these charm baryons to the $\rm D^0$ meson are presented in Fig.~\ref{fig:sigmac_xic}, compared to predictions from the same models that try to describe the $\rm \Lambda_c^+ / D^0$ ratio~\cite{Skands:2014pea,Christiansen:2015yqa,Minissale:2020bif,He:2019tik}. The predictions tuned on measurements in $\rm e^{+}e^{-}$ collisions~\cite{Skands:2014pea} clearly underestimate these heavier charm baryon-to-meson ratios, providing further evidence that different processes are involved in charm hadronisation for elementary and hadronic collisions. Model predictions based on colour reconnections beyond leading-colour approximation~\cite{Christiansen:2015yqa} or on statistical hadronisation with additional baryon states~\cite{He:2019tik} are both in agreement with the $\rm \Sigma_c^{0,+,++}(2455)$ ratio measurement. For the $\rm \Xi_c^{0,+} / D^0$ ratio~\cite{Acharya:2021vjp}, all models except the Catania model~\cite{Minissale:2020bif} (relying on hadronisation via recombination after the formation of a colour-deconfined state of matter) are underpredicting the measured ratios, hinting to an even stronger enhancement for charm-strange baryons.

\section{Conclusion}

In this contribution, the most recent results on charm production and hadronisation in pp and Pb--Pb collisions measured by the ALICE Collaboration were presented. While for the $\rm D$-meson sector, calculations based on the factorisation approach tuned on previous measurements from $\rm e^+e^-$ colliders describe well the data, the production of charm baryons is significantly enhanced with respect to such predictions. This puzzle triggered several new measurements on the production of $\rm \Lambda_c^+$ versus event multiplicity in different collision systems and on the production of heavier charm-baryon states, as well as developments on the theoretical side to explain the observations. Despite these efforts, the picture of charm hadronisation is not yet complete, and further observables and extra decay channels/baryons will be studied in the near future with data from LHC Run 3.


\section*{References}

\begin{thebibliography}{99}
\bibitem{Andronic:2015wma} A.~Andronic \textit{et al.} Eur. Phys. J. C \textbf{76} (2016) no.3, 107 [arXiv:1506.03981 [nucl-ex]].
\bibitem{Acharya:2017kfy} S.~Acharya \textit{et al.} [ALICE], JHEP \textbf{04} (2018), 108 [arXiv:1712.09581 [nucl-ex]].
\bibitem{Acharya:2019mgn} S.~Acharya \textit{et al.} [ALICE], Eur. Phys. J. C \textbf{79} (2019) no.5, 388 [arXiv:1901.07979 [nucl-ex]].
\bibitem{Cacciari:2012ny} M.~Cacciari \textit{et al.} JHEP \textbf{10} (2012), 137 [arXiv:1205.6344 [hep-ph]].
\bibitem{Acharya:2021cqv} S.~Acharya \textit{et al.} [ALICE], [arXiv:2102.13601 [nucl-ex]].
\bibitem{Acharya:2020uqi} S.~Acharya \textit{et al.} [ALICE], [arXiv:2011.06078 [nucl-ex]].
\bibitem{Gladilin:2014tba} L.~Gladilin, Eur. Phys. J. C \textbf{75} (2015) no.1, 19 [arXiv:1404.3888 [hep-ex]].
\bibitem{Skands:2014pea} P.~Skands \textit{et al.} Eur. Phys. J. C \textbf{74} (2014) no.8, 3024 [arXiv:1404.5630 [hep-ph]].
\bibitem{Bahr:2008pv} M.~Bahr \textit{et al.} Eur. Phys. J. C \textbf{58} (2008), 639-707 [arXiv:0803.0883 [hep-ph]].
\bibitem{Christiansen:2015yqa} J.~R.~Christiansen and P.~Z.~Skands, JHEP \textbf{08} (2015), 003 [arXiv:1505.01681 [hep-ph]].
\bibitem{Minissale:2020bif} V.~Minissale \textit{et al.} [arXiv:2012.12001 [hep-ph]].
\bibitem{He:2019tik} M.~He and R.~Rapp, Phys. Lett. B \textbf{795} (2019), 117-121 [arXiv:1902.08889 [nucl-th]].
\bibitem{Acharya:2021vjp} S.~Acharya \textit{et al.} [ALICE], [arXiv:2105.05187 [nucl-ex]].
\bibitem{Acharya:2021set} S.~Acharya \textit{et al.} [ALICE], [arXiv:2105.06335 [nucl-ex]].
\end{thebibliography}
\end{document}